\documentclass[twocolumn,showpacs,prl,preprintnumbers,amsmath,amssymb]{revtex4}
\usepackage{amsmath}
\usepackage{graphicx}

\newcommand{\beq}{\begin{equation}}
\newcommand{\eeq}{\end{equation}}
\newcommand{\beqa}{\begin{eqnarray}}
\newcommand{\eeqa}{\end{eqnarray}}

\def\ket#1{\mathinner{|{#1}\rangle}}

\newcommand{\pd}[2]{\frac{\partial #1}{\partial #2}}

\newcommand{\roth}{\rho_{13}}
\newcommand{\rof}{\rho_{14}}
\newcommand{\rtth}{\rho_{23}}
\newcommand{\rtf}{\rho_{24}}

\newcommand{\oma}{$\Omega_a$ }
\newcommand{\omb}{$\Omega_b$ }
\newcommand{\omc}{$\Omega_c$ }
\newcommand{\omd}{$\Omega_d$ }

\def\josab#1{{ J. Opt.\ Soc.\ Am.\ B\/} {\bf#1}}

\def\pr#1{{ Phys.\ Rev. } {\bf#1}}
\def\pra#1{{ Phys.\ Rev. A\/} {\bf#1}}

\def\prl#1{{ Phys.\ Rev.\ Lett.} {\bf#1}}

\begin{document}

\title{Pulse Areas in Multi-Soliton Propagation}

\author{Elizabeth Groves}
\email{egroves@pas.rochester.edu}

\author{B.D. Clader}

\author{J.\ H.\ Eberly}

\affiliation{ Rochester Theory Center \\
and \\
Department of Physics \& Astronomy\\
University of Rochester, Rochester, New York 14627}
\pacs{42.65.Tg, 42.50.Nn, 42.50.Gy, 45.50.Md}
\date{\today}
\begin{abstract}  The prospect of self-consistent propagation of more than two pulses contemporaneously through multi-resonant media raises open questions: whether soliton solutions exist, and whether a useful generalization of two-level pulse Area can be found. We answer these questions positively for the case of four pulses interacting in combined $V$ and $\Lambda$ fashion with an idealized pair of atomic D-lines.

\end{abstract}

\maketitle

%\section*{Introduction}

The fully coherent propagation of light pulses in absorbers with a dominant resonant transition was given a new foundation, superseding Lorentzian theory, by the work of McCall and Hahn \cite{McCall-Hahn} in the late 1960s.  Their theory predicted stable solitonic pulses and introduced a new concept, the dimensionless quantum pulse ``Area", $\theta(z,t)$, as the time integral of the pulse envelope ${\cal E}(z,t)$:
\beqa \label{Area}
\theta(z,t) & \equiv & \frac{2d}{\hbar}\int_{-\infty}^t {\cal E}(z,t') dt' \nonumber\\
& = & \int_{-\infty}^t \Omega(z,t') dt', 
\eeqa
where  
\beq \label{RabiFreq}
\Omega(z,t) \equiv  \frac{2d}{\hbar}{\cal E}(z,t)
\eeq
is the well-known Rabi frequency and $d$ is the dipole transition matrix element for the resonant transition. The $\hbar$ in its denominator indicates that the Rabi frequency has no classical $\hbar \to 0$ correspondence limit. The McCall-Hahn ``Area Theorem" establishes that $\theta = 2n\pi$ identifies values for Area that are stable in the propagation of pulses whose duration is much shorter than relevant homogeneous relaxation times such as the spontaneous lifetime \cite{Allen-Eberly}.  

\begin{figure}[!b]
\centerline{\includegraphics[width=8cm]{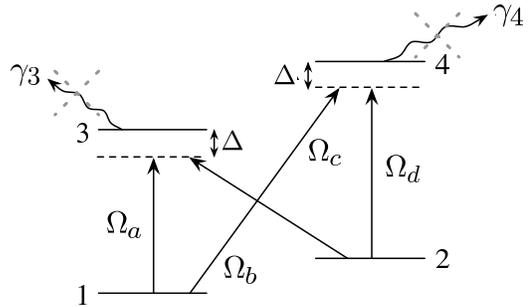}}
\caption{\label{f.system}A four-level atom interacting with four separate lasers with equal detunings, $\Delta$, and four laser pulses indicated by their Rabi frequencies.  In the short-pulse regime, we can neglect relaxation processes, as indicated by the crosses in the rates $\gamma_3$ and $\gamma_4$. The sketch can be interpreted as showing  idealized alkali D line transitions, where levels 3 and 4 are $P_{1/2}$ and $P_{3/2}$ manifolds, and levels 1 and 2 correspond to F=1 and F=2 hyperfine levels of $S_{1/2}$.}
\end{figure}

The propagation of several fields simultaneously through multi-level media with several resonant transitions opens two questions: whether stable soliton solutions still exist for any such cases and whether pulse Area remains a key parameter. Here we examine these questions by extending recent work \cite{Clader-Eberly07, Clader-Eberly08} on three-level media to more complex four-level media that can be viewed as having either double $\Lambda$ or double $V$ form for their levels and resonances, as shown in Fig.\ \ref{f.system}.

In the context of Electromagnetically Induced Transparency (EIT) \cite{Harris} pulse propagation in three-level media has been studied extensively \cite{Harris93, Harris94, Eberly-etal, Kasapi-etal95}, and for multi-level generalizations of EIT \cite{Schmidt, Paspalakis-Knight, McGloin, Deng-Payne}. Transparency in these systems is usually achieved by having a strong quasi-cw field produce a steady-state dressing of the medium on one resonance, producing a very narrow spectral window for essentially lossless propagation of a much weaker pulse on the other transition.  In the opposite regime, where the pulses are wideband and short, the evolution of both pulses is coherently dynamic, leading to soliton formation. This situation has been examined for particular multi-level media \cite{Konopnicki-Eberly, Basharov-Maimistov, Bolshov, Hioe-GellMann, Hioe-Grobe, Rahman-Eberly, Park-Shin, Rahman, Eberly-Kozlov-DAT, Rybin-Vadeiko, Huang-Hang-Deng, Clader-Eberly07, Clader-Eberly08}, but the role of pulse Area has been largely overlooked.  

Under the condition of pulse-matching \cite{Hioe-Eberly84, Harris93}, simulton pulses \cite{Konopnicki-Eberly} are known to obey a two-pulse Area Theorem \cite{Rahman, Rahman-Eberly} in $V$-type media and multi-level generalizations exist for certain systems \cite{Hioe-GellMann}.  The recent result of Clader and Eberly \cite{Clader-Eberly08} gave an exact analytic solution for $\Lambda$-type media in which the pulses are temporally matched but have different spatial propagation.  Eberly and Kozlov have shown that the propagation of unmatched two-photon-resonant pulse pairs in $\Lambda$-type media depends on the so-called dark Rabi frequency \cite{Fleischhauer-Manka} and obeys a dark Area Theorem \cite{Eberly-Kozlov-DAT}.  

In this work, we examine four-pulse propagation under conditions suggested by Fig.\ \ref{f.system}, such that the durations of all pulses are short enough to neglect homogeneous relaxation.  Hioe has shown \cite{Hioe-GellMann} that this level scheme permits simulton solutions (multiple pulses travelling with the same group velocity).  Here we employ the B\"acklund Transformation method \cite{Lamb} of Park and Shin \cite{Park-Shin}. It is robust enough to accommodate both inhomogeneous broadening, which we take into account, and more than a single pulse velocity.  As is evident in Fig.\ \ref{f.system}, atoms in the four-level configuration can be viewed as a double $\Lambda$- or double $V$-type medium and our solution has attributes of known $\Lambda$ and  $V$ solutions.  We have discovered that pulse Area remains a parameter important to the evolution of our system and that three ``total" pulse Areas are constant.
%\section*{Physical System}

For the four short optical pulses, we assume that the laser fields are linearly polarized and can be written in envelope-carrier wave form as  $E_a(x,t) = \mathcal{E}_a(x,t)e^{(i k_a x - \omega_a t)} + c.c.,$ where $\mathcal{E}_a(x,t)$ is the slowly-varying field envelope function, $k_a$ is the wave number and $\omega_a$ is the field frequency addressing the 1-3 transition labelled $\Omega_a.$  Similar notation applies to the other laser fields and each is assumed to address a single atomic transition.  Under the Rotating Wave Approximation (RWA), the Hamiltonian for this atom-laser system can be written (in the $\ket{1},\ket{2},\ket{3},\ket{4}$ basis indicated in Fig.\ \ref{f.system}) as

\beq
H = 
-\hbar\begin{pmatrix}
0& 0 &\frac{1}{2}\Omega_a &\frac{1}{2}\Omega_b \\
0& 0 &\frac{1}{2}\Omega_c &\frac{1}{2}\Omega_d \\
\frac{1}{2}\Omega^*_a & \frac{1}{2}\Omega^*_c & \Delta &0 \\
\frac{1}{2}\Omega^*_b & \frac{1}{2}\Omega^*_d &0 & \Delta
\end{pmatrix},
\eeq
where the Rabi frequency $\Omega_a$ governs transitions between levels 1 and 3 through the atomic dipole moment operator $d_a$ associated with the $a$ transition, and the other Rabi frequencies are similarly defined.  

Transitions between levels $1$ and $2$, as well as between levels $3$ and $4$, are assumed to be dipole forbidden due to the parity of the levels.  We require the detuning to be the same for all transitions: $\Delta = \omega_3 - (\omega_1+\omega_a) = \omega_3-(\omega_2+\omega_c) = \omega_4 - (\omega_1+\omega_b) = \omega_4 - (\omega_2+\omega_d)$.  As indicated in Fig.\ \ref{f.system}, we ignore homogeneous relaxation rates, such as $\gamma_3$ and $\gamma_4$.  The atomic density matrix then satisfies the von Neumann equation    
\beq
i\hbar\pd{\rho}{t} = [H,\rho].
\eeq

The propagation of each field is governed by Maxwell's equations which, under the Slowly-Varying Envelope Approximation (SVEA), become:
 \begin{subequations}\label{e.Max: 1}
 \begin{eqnarray}
 \pd{\Omega_a}{Z} = -i \mu_a\left< \roth\right>, & & \label{e.Max: 1A}
\pd{\Omega_b}{Z} =  -i \mu_b\left< \rof\right>,\\ 
\pd{\Omega_c}{Z} = -i \mu_c\left< \rtth \right>,& & \label{e.Max: 1B}
\pd{\Omega_d}{Z} =-i \mu_d\left<  \rtf \right>,
\end{eqnarray}
\end{subequations}
where we use the retarded coordinates $T= t-x/c$ and $Z = x/c$.  The atom-field coupling parameter is $\mu_a = \mathcal{N}d_a^2 \omega_a/\hbar\epsilon_0$, where $\mathcal{N}$ is the density of atoms.  Similar notation applies to $\mu_b$, $\mu_c$, and $\mu_d$.  The brackets denote an average over all atoms to account for inhomogeneous broadening.  We assume an inhomogeneous lifetime $T^*_2$ and line-center tuning.  The  distribution function is then
$F(\Delta) = ({T_2^*}/{\sqrt{2\pi}})exp[-(\Delta T_2^*)^2/2]$, and the averaging is performed by integrating over the detunings.  
The form of the von-Neumann equation remains the same in the retarded coordinates but the derivative must be taken with respect to the retarded time. 

We apply the  B\"acklund Transformation method of \cite{Clader-Eberly07, Park-Shin} to our four-level system by first recasting Eqn.\ \eqref{e.Max: 1} in terms of the Hamiltonian.  This is easily done by introducing a constant matrix 
 \begin{equation}
W = i
\begin{pmatrix}
0& 0 &0 &0 \\
0& 0 &0 &0 \\
0& 0 &1 &0 \\
0& 0 &0 &1
\end{pmatrix}
\end{equation}
so that the slowly-varying Maxwell equations can be written in commutator form:
 \beq\label{Hz}
 \pd{H}{Z} =-\frac{\hbar \mu}{2} [W,\rho].
 \eeq
Note that we have required the atom-field coupling for each transition to be the same:  $\mu\equiv\mu_a =\mu_b=\mu_c=\mu_d$.
%The B.T. method of \cite{Clader-Eberly} exploited the fact that Maxwell's slowly varying equation could be rewritten in terms of the Hamiltonian and the density matrix, $\rho$ by introducing a constant matrix $W$ such that 

Written in this way, Eqns.\ \eqref{e.Max: 1} and \eqref{Hz} differ from the $\Lambda$-system case only in the definition of $W$.  The solution method outlined by Clader and Eberly, based on the B\"acklund Transformation of Park and Shin \cite{Park-Shin}, does not employ the explicit form of $W$ and can thus be used to generate solutions to our system.  

%\section*{Analytic Solutions}
Given any ``seed" soliton solution to a set of non-linear partial differential equations, the B\"acklund Transformation generates a higher-order soliton solution from it.  An obvious exact solution to Eqns.\ \eqref{e.Max: 1} and \eqref{Hz} is given by zero applied fields:
\beq
\Omega_a(T,Z) = \Omega_b(T,Z) = \Omega_c(T,Z) = \Omega_d(T,Z) = 0, \nonumber
\eeq
and constant level populations, as specified by this mixed state density matrix:
\beq
\rho(T,Z) = 
\rho(0,0) =
\begin{pmatrix}
|\alpha|^2 & 0 &0 &0 \\
0& |\beta|^2 &0 &0 \\
0& 0 &0 &0 \\
0& 0 &0 &0
\end{pmatrix},
\eeq
where $|\alpha|^2 + |\beta|^2 =1$.  To be specific, we will assume in the present discussion that $|\alpha|^2 \geq |\beta|^2$.
We have applied the B\"acklund Transformation method to the seed solution above, which leads to exact solutions for the Rabi frequencies and atomic density matrix elements.  We will focus on the Rabi frequencies  and will not list the density matrix elements here.  The Rabi frequencies are:    
\begin{subequations} \label{e.pulses}
\beqa
\Omega_a(Z,T) = \sin{u}\sin{\phi(Z)}\Omega(Z,T), \\
\Omega_b(Z,T) = \cos u \sin \phi(Z)\Omega(Z,T),\\
\Omega_c(Z,T) = \sin u \cos \phi(Z)\Omega(Z,T),\\
\Omega_d(Z,T) = \cos u \cos \phi(Z) \Omega(Z,T), 
\eeqa
\end{subequations}
where the fundamental Rabi frequency $\Omega$ has the complicated form:
\beq
\Omega(Z,T)=   \frac{ (4/\tau)\sqrt{1+\exp{[2(\alpha^2-\beta^2)\kappa Z]}}}{2 \cosh\left[\alpha^2\kappa Z  - \frac{T}{\tau}\right]  +  \exp{\left[(\alpha^2-2\beta^2)\kappa Z + \frac{T}{\tau}\right]}} \nonumber
\eeq
and $\tan \phi(Z) = \exp[ (\beta^2-\alpha^2)\kappa Z]$.

In the expressions above, $\tau$ is the nominal pulse width and the length scale is set by $\kappa$ defined as the inhomogeneous average  $\kappa = ({\mu}/{2\tau}) \langle {1}/({\Delta^2 + {1}/{\tau}^2})\rangle$.
The parameter $u$ is a constant of integration that specifies the initial conditions of the pulses. For $\Lambda$-type media, this constant is of little physical interest because it merely shifts the origin in the spatially infinite medium.  In the four-level system, however, $u$ has a natural interpretation and allows freedom in the choice of the relative pulse amplitudes.  

The pulse propagation behavior is difficult to infer from the complicated formulas above.  However, plots show that the evolution can be broken down into three distinct regimes: regime I, where $Z$ and $T$ are large and negative; regime II, where $Z,T \sim 0$; regime III where $Z$ and $T$ are large and positive.

In the asymptotic limit of regime I, where $-\kappa Z, -T/\tau \gg 1$, pulses \oma and \omb have $sech$ form, with Areas equal to $2\pi\sin u$ and $2\pi\cos u$, respectively, and pulses \omc and \omd approach zero amplitude. In regime II the complicated 4-level interaction process begins to convert pulses \oma and \omb into pulses \omc and $\Omega_d,$ which eventually become $sech$ pulses in regime III, where $\kappa Z, T/\tau \gg 1$. They asymptotically approach Areas equal to $2\pi\sin u$ and $2\pi\cos u$, respectively. The input and output pulse pairs are not identical, however, if $\alpha \neq \beta$. Then the input group velocity $v_g = c(1+\alpha^2\kappa\tau)^{-1}$ shared by \oma and \omb differs from the output group velocity $v_g = c(1+\beta^2\kappa\tau)^{-1}$ shared by \omc and $\Omega_d,$ and the output velocity is greater under our assumption that $\alpha > \beta$. The conversion from pulses \oma and \omb to pulses \omc and \omd is a feature not present in the pure ``simulton" pulses of Hioe \cite{Hioe-GellMann}, which all travel with the same group velocity because $\alpha = \beta$ is imposed in that case.

%\section*{Numerical Solutions}
We need to probe the stability of these solutions.  That is, we need to know if these analytic solutions to idealized medium and pulse conditions can provide useful experimental insights.  Thus, we have employed numerical methods to calculate the expected behavior of non-ideal input pulses in finite media.  

\begin{figure}[!h]
\centerline{\includegraphics[width=8.0cm]{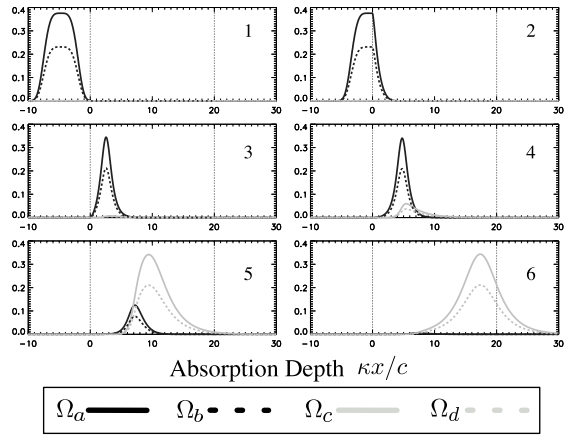}}
\caption{\label{f.NumSol} 
Snapshots of time-matched pulses represented by their Rabi frequencies generated from numerical simulations.  The position of the medium is represented by the vertical lines.  All pulses are initially of square-Gaussian shape with initial areas approximately those of regime I .  We have chosen $\theta_a = 1.4\pi$, $\theta_b=0.9\pi $, $\theta_c = 0.002\pi$, and $\theta_d=0.001\pi$ so that $\theta_T$ is less than $2\pi$ in Frame 1. }
\end{figure}

A particular example is shown in Fig.\ \ref{f.NumSol}.  The simulation was performed with square-Gaussian input pulses (i.e., having quartic rather than quadratic exponents) with total Area $\theta_T$  less than the predicted stable solitonic value of $2\pi$.  In contrast to the analytic conditions of an infinite medium, i.e., zero amplitude for pulses \omc and \omd in the asymptotic input regime, we alloted a small initial amplitude to pulses \omc and \omd before they enter the medium.  

%The three-regime behavior predicted analytically is clearly evident in the numerical solutions.  Upon entry into the medium, pulses \oma and \omb are quickly reshaped toward the preferred solitonic $sech$ form (Fig.\ \ref{f.NumSol}, Frame 2).  With further penetration into the medium, the weak pulses \omc and \omd are amplified in the pulse transfer regime and \oma and \omb are depleted (Fig.\ \ref{f.NumSol}, Frames 3-5).  Finally (Fig.\ \ref{f.NumSol}, Frame 6), \omc and \omd propagate without attenuation through the remainder of the medium. Their greater width along the coordinate axis is a consequence of their greater group velocity.
Upon entry into the medium, pulses \oma and \omb are quickly reshaped toward the preferred solitonic $sech$ form (Fig.\ \ref{f.NumSol}, Frame 2).  The $sech$ pulses travel at a reduced group velocity in the medium, which results in their reduced width along the coordinate axis (Fig.\ \ref{f.NumSol}, Frame 3).  With further penetration into the medium, the weak pulses \omc and \omd are amplified in the pulse transfer regime as \oma and \omb are depleted (Fig.\ \ref{f.NumSol}, Frames 4,5).  Finally (Fig.\ \ref{f.NumSol}, Frame 6), \omc and \omd are of stable $sech$ form and propagate without attenuation through the remainder of the medium.  The greater width of the ``output" pulses \omc and $\Omega_d,$ in comparison with \oma and $\Omega_b$, along the coordinate axis is a consequence of their greater group velocity in the medium.  

The parameter $u$ determines the maximum amplitudes of the pulses and clearly shows that the maximum amplitude of \omc is determined by the maximum amplitude of \oma and, similarly, that the growth of \omd is limited by $\Omega_b.$  This behavior can be best understood when viewed in the context of pulse Area, as follows.

%\section*{Pulse Area}
Pulse transfer processes were previously observed for $\Lambda$-type media by Clader and Eberly \cite{Clader-Eberly07, Clader-Eberly08}.  They found that the individual pulse Areas, defined by $\theta(Z) = \int^{\infty}_{-\infty} \Omega(Z,T) dT$, were constant in the asymptotic regimes I and III and that, surprisingly, a ``total" pulse Area was constant throughout the entire medium.  We have found that similar behavior occurs in a four-level medium.  We can calculate the Area of each pulse from Eqn.\ \eqref{e.pulses} and find
\begin{subequations} \label{e.areas: 1}
\begin{align} 
\theta_a(Z) &=2\pi \sin u \sin \phi(Z), \\
\theta_b(Z) &=  2\pi \cos u \sin \phi(Z), \\
 \theta_c(Z) &=2\pi \sin u \cos \phi(Z), \\
 \theta_d(Z) &=2\pi \cos u \cos \phi(Z).
\end{align}
\end{subequations}

We note that particular combinations of these Areas are constant throughout propagation.  In particular, we find
\begin{subequations} \label{e.totalareas: 1}
\begin{align}
\theta_1(Z) &\equiv \sqrt{|\theta_a(Z)|^2+|\theta_c(Z)|^2} = 2\pi \sin u \label{e.totalareas: 1A}, \\
\theta_2(Z) &\equiv \sqrt{|\theta_b(Z)|^2+|\theta_d(Z)|^2} = 2\pi \cos u \label{e.totalareas: 2B}, \\ 
\theta_T(Z) &\equiv \sqrt{ |\theta_1(Z)|^2 + |\theta_2(Z)|^2} = 2\pi.  \end{align}
\end{subequations}
As in the $\Lambda$-system, the ``total" pulse Area $\theta_T$ is exactly $2\pi$ throughout propagation, for any value of $Z$.  In addition, the four-level system has two more relations, Eqns.\ \eqref{e.totalareas: 1A} and \eqref{e.totalareas: 2B}, restricting the pulse Areas.  These clearly show that amplification of \omc is a consequence of the depletion of \oma and that, similarly, the Area of \omd grows at the expense of the Area of $\Omega_b.$   

\begin{figure}[!t]
\centerline{\includegraphics[width=8cm]{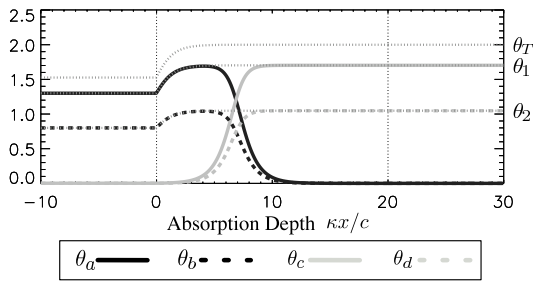}}
\caption{\label{f.NumArea} Pulse areas in units of $\pi$ as a function of propagation distance in the medium, indicated by the vertical lines, from numerical simulation.  The dotted lines show the ``total" pulse areas $\theta_1$, $\theta_2$, and $\theta_T$, as indicated on the right side of the figure.}
\end{figure}

To check these predictions derived for ideal pulses and media we show the pulse Areas corresponding to the pulse evolution in Fig.\ \ref{f.NumSol}, also computed numerically. They are plotted in Fig.\ \ref{f.NumArea}. One easily sees that the total pulse Areas grow to the analytically predicted values.

%\section*{Conclusions}
In this note we have presented new exact analytic solutions to a four-pulse propagation problem, extending previous simulton results into a regime that can be identified as four-wave mixing of solitons.  The solutions have been shown to combine elements of known solutions for $V$- and $\Lambda$-type media.  The three regime pulse-transfer behavior of $\Lambda$ systems identified in \cite{Clader-Eberly07} is again present.  We have shown that pulse Area remains a powerful nonlinear constraint on the evolution of pulses in four-level media, and more general combinations of the pulse Areas are stable: three ``total" pulse Areas remaining constant throughout the medium.  Numerical simulations have been used to confirm the utility of our analytic solutions under non-ideal conditions for which such solutions are not available.  

Acknowledgements: We acknowledge helpful cooperation with Prof. Q.-H. Park, financial support from NSF Grant PHY-0601804, and a Horton Fellowship awarded E. Groves.


\begin{thebibliography}{99}
\bibitem{McCall-Hahn} S. L. McCall and E. L. Hahn, Phys. Rev. Lett. {\bf 18}, 908 (1967); and \pr {183}, 457 (1969)

\bibitem{Allen-Eberly} For an extended discussion, see L. Allen and J.H. Eberly, {\em Optical Resonance and Two-Level Atoms}, (Dover, 1987).
 
\bibitem{Clader-Eberly07} B. D. Clader and J. H. Eberly, \pra{76}, 053812 (2007)

\bibitem{Clader-Eberly08} B. D. Clader and J. H. Eberly, \pra{78}, 033803 (2008)

\bibitem{Harris} S. E. Harris, Phys. Today {\bf 50}, 36 (1997)

\bibitem{Harris93} S.E. Harris, \prl {70}, 552-555 (1993) 

\bibitem{Harris94} S.E. Harris, \prl {72}, 52 (1994) 

\bibitem{Eberly-etal}  J.H. Eberly, M.L. Pons and H.R. Haq, \prl {72}, 56 (1994)

\bibitem{Kasapi-etal95} A. Kasapi, M. Jain, G.Y. Yin, et al. \prl {74}, 2447 (1995) 

\bibitem{Schmidt} H. Schmidt and A. Imamoglu, Opt. Lett. {\bf 21}, 1936 (1996)
\bibitem{McGloin} D. McGloin, D. J. Fulton, and M. H. Dunne, Opt. Commun. {\bf 190}, 221 (2001) 
\bibitem{Paspalakis-Knight} E. Paspalakis and P. L. Knight, \pra{66}, 015802 (2002)
\bibitem{Deng-Payne} L. Deng, M. G. Payne, Phys. Rev. Lett {\bf 91}, 243902 (2003)
%*********************************
%solitons in multilevel media
\bibitem{Konopnicki-Eberly} M. J. Konopnicki and J. H. Eberly, Phys. Rev. A {\bf 24}, 2567 (1981)

%\bibitem{Maimistov} A. I. Maimistov, Sov. J. Quantum Electron. {\bf 14}, (1984)
%\bibitem{Basharov
\bibitem{Basharov-Maimistov} A. M. Basharov and A. I. Maimistov, Sov. Phys. JETP {\bf 67}, 2426 (1988)
\bibitem{Bolshov} L. A. Bol'shov and V. V. Likhanskii, Sov. J. Quantum Electron. {\bf 15}, 889 (1985)
\bibitem{Hioe-GellMann} F. T. Hioe, \josab{6}, 1245 (1989)
\bibitem{Hioe-Grobe} F. T. Hioe and R. Grobe, \prl{73}, 2559 (1994)
\bibitem{Rahman-Eberly} A. Rahman and J. H. Eberly, Phys. Rev. A {\bf 58}, R805 (1998)
\bibitem{Park-Shin} Q-Han Park and H. J. Shin, Phys. Rev. A {\bf 57}, 4643 (1998)
\bibitem{Rahman} A. Rahman, Phys. Rev. A. {\bf 60}, 4187 (1999)
\bibitem{Eberly-Kozlov-DAT} J. H. Eberly and V. V. Kozlov, Phys. Rev. Lett. {\bf 88}, 243604 (2002)
\bibitem{Rybin-Vadeiko} A. V. Rybin and I. P Vadeiko, \josab{6}, 416 (2004)
\bibitem{Huang-Hang-Deng} G. Huang, C. Hang, and L. Deng, Eur. Phys. J. D. {\bf 40}, 437 (2006) 

\bibitem{Hioe-Eberly84} F.T. Hioe and J.H. Eberly, \pra{29}, 1164 (1984)

\bibitem{Fleischhauer-Manka} M. Fleischhauer and A. S. Manka, \pra{54} 794 (1996)

%**********************************
\bibitem{Lamb} G. L. Lamb, Jr. {\it Elements of Soliton Theory} (Wiley, New York, 1980)
%Numerical lambda:
%\bibitem{Deng-Payne-Garrett} L. Deng, M. G. Payne, and W. R. Garrett, Opt. Commun. {\bf 242}, 641 (2004)
%\bibitem{Clader-Raman} B. D. Clader and J. H. Eberly, in {\it Proceedings of the International Symposium on Quantum Optics, Ahmedabad}, edited by J. Banerji, P. K. Panigrahi, and R. P. Singh (Macmillan of India, Delhi, 2007), p. 3.
%G. S. Agarwal and J. H. Eberly, \pra{61} 013404 (1994);
%R. Grobe, F. T. Hioe and J. H. Eberly \prl{73}, 3183 (1994);
%J. H. Eberly, Quantum Semiclassic. Opt. {\bf 7} 373 (1995);
%G. Huang, C. Hang, and L. Deng, Eur. Phys. J. D. {\bf 40}, 437 (2006)

\end{thebibliography}
\end{document}